\pdfoutput=1
\documentclass[aps,prl,floatfix,twocolumn,reprint]{revtex4}
\usepackage{eurosym}
\usepackage{amsbsy}
\usepackage{latexsym,epsfig,graphicx}
\usepackage{dcolumn}
\usepackage{graphicx}
\usepackage{subfigure}
\usepackage{comment}
\usepackage{color}
\usepackage{bm}
\usepackage{mathrsfs}
\usepackage{amssymb}
\usepackage{amsfonts}
\usepackage{amsmath}
\usepackage{xspace}
\usepackage{epstopdf}
\usepackage{tabularx}
\usepackage{longtable}
\usepackage[colorlinks=true, letterpaper=true, pdfstartview=FitV, linkcolor=red, citecolor=blue, urlcolor=blue]{hyperref}
\usepackage[normalem]{ulem}
\usepackage[version=4]{mhchem}

\begin{document}

\title{Spin-1 Topological Monopoles in Parameter Space of Ultracold Atoms}
\author{Haiping Hu}
\affiliation{Department of Physics, The University of Texas at Dallas, Richardson, Texas 75080, USA}
\author{Chuanwei Zhang}
\email{chuanwei.zhang@utdallas.edu}
\affiliation{Department of Physics, The University of Texas at Dallas, Richardson, Texas 75080, USA}
\date{\today}

\begin{abstract}
Magnetic monopole, a hypothetical elementary particle with isolated magnetic pole, is crucial for the quantization of electric charge. In recent years, analogues of magnetic monopoles, represented by topological defects in parameter spaces, have been studied in a wide range of physical systems. These works mainly focused on Abelian Dirac monopoles in spin-1/2 or non-Abelian Yang monopoles in spin-3/2 systems. Here we propose to realize three types of spin-1 topological monopoles and study their geometric properties using the parameter space formed by three hyperfine states of ultracold atoms coupled by radio-frequency fields. These spin-1 monopoles, characterized by different monopole charges, possess distinct Berry curvature fields and spin textures, which are directly measurable in experiments. The topological phase transitions between different monopoles are accompanied by the emergence of spin \textquotedblleft vortex\textquotedblright , and can be intuitively visualized using Majorana's stellar representation. We show how to determine the Berry curvature, hence the geometric phase and monopole charge from dynamical effects. Our scheme provides a simple and highly tunable platform for observing and manipulating spin-1 topological monopoles, paving the way for exploring new topological quantum matter.
\end{abstract}

\maketitle

\emph{Introduction}\textit{.---}In 1931, Dirac proposed the quantum theory of magnetic charge, in which the existence of magnetic monopole leads to the quantization of electric charge \cite{dirac}. The magnetic monopole carrying a net magnetic charge $\hbar/2e$, is considered as the source for induced magnetic field $\bm{B}$ satisfying Gauss' law $\oint_{\bm{\mathcal{S}}}\bm{B}\cdot d\bm{S}=nh/e$, where $n$ counts the number of magnetic charges enclosed by a two-dimensional (2D) integral manifold $\bm{\mathcal{S}}$. Although no direct experimental evidence for magnetic (Dirac) monopoles has been reported so far, analogues of magnetic monopoles have been found in various physical systems \cite{monocond1,monocond2,monocond3,monocond4,monocond5,monocond6,monoexp1,monoexp2}. In such Dirac-like monopoles, the monopole charge is defined as the topological invariant
\begin{equation*}
\mathcal{C}=\frac{1}{2\pi }\oint_{\bm{\mathcal{S}}}\bm{\Omega}\cdot d\bm{S},~~~\bm{\Omega }=\bm{\nabla}_{\bm{R}}\times \langle \psi |i\bm{\nabla}_{\bm{R}}|\psi \rangle ,
\end{equation*}
i.e., the first Chern number, where $\bm{R}$ represents an extended state space (e.g., position, momentum, or certain parameters) and the Berry curvature $\bm{\Omega}$ corresponds to the effective magnetic field. In many condensed matter materials, such as recently discovered Dirac \cite{dirac1,dirac2,dirac3,dirac4} and Weyl semimetals \cite{weyl1,weyl2,weyl3,weyl4,weyl5,weyl6,weyl7,weyl8,weyl9,weyl10,weyl11,weyl12,weyl13,Gong2011,Yong2015}, topological monopoles usually represent Berry curvature singularities in the momentum space with energy level degeneracy. The monopole charges are topologically protected against small perturbations and their changes indicate topological phase transitions.

While Dirac-like monopoles have been broadly investigated in both theory and experiment for spin-1/2 systems because of their significance for characterizing topological quantum matter and realizing geometric quantum computation \cite{Duan2001,Leibfried,Nielsen}, the study of topological monopoles in higher spin systems started to attract attentions in recent years, with important experimental progress such as the observation of non-Abelian Yang monopoles \cite{yang} in a degenerate state space for a spin-3/2 atomic gas \cite{yangexp}. In this context, spin-1 topological monopoles may be of special interests because, unlike spin-1/2 and 3/2, the underlying Hamiltonian cannot be written as direct products of Pauli matrices, and naturally contains spin quadrupole tensors. For any $3\times 3$ Hamiltonian, these spin vectors and tensors are equivalent to the so-called Gell-Mann matrices, which form a basis of the SU(3) algebra. The three-component quantum state cannot be simply mapped onto a Bloch sphere,
therefore the geometric phase and monopole charge cannot be simply determined by the solid angle and winding number on the sphere \cite{monoexp1,monoexp2}. In the momentum space, spin-1 topological monopoles correspond to triply-degenerate band-touching points, which have been studied in solid states \cite{tf1} and cold atoms \cite{tf2,tf3} recently, but their experimental realization is still elusive.

In this paper, we propose that ultracold atoms in the parameter space formed by radio-frequency (rf) couplings between three hyperfine atomic ground states provide a simple and highly tunable platform for studying exotic topological phases in spin-1 systems, in particular, spin-1 topological monopoles. Three types of spin-1 monopoles and their topological phase transitions are characterized by the emergence of spin vortices, which change the spin textures in the parameter space and can be directly probed in experiments.
These monopoles can also be visualized using Majorana's stellar representation (MSR) \cite{majorana1} on the state space. With this geometric representation, different monopoles yield topologically distinct trajectories of Majorana stars on the Bloch sphere. Finally, a dynamical protocol \cite{pnas} is proposed to measure the Berry curvature in the parameter space, which determines the geometric phase \cite{berry} associated with an adiabatic evolution path as well as the topological monopole charge. Our scheme can be easily generalized to larger spin systems, providing a general platform for studying exotic topological quantum matter using the parameter space of ultracold atoms.

\begin{figure}[t]
\centering
\includegraphics[width=3.3in]{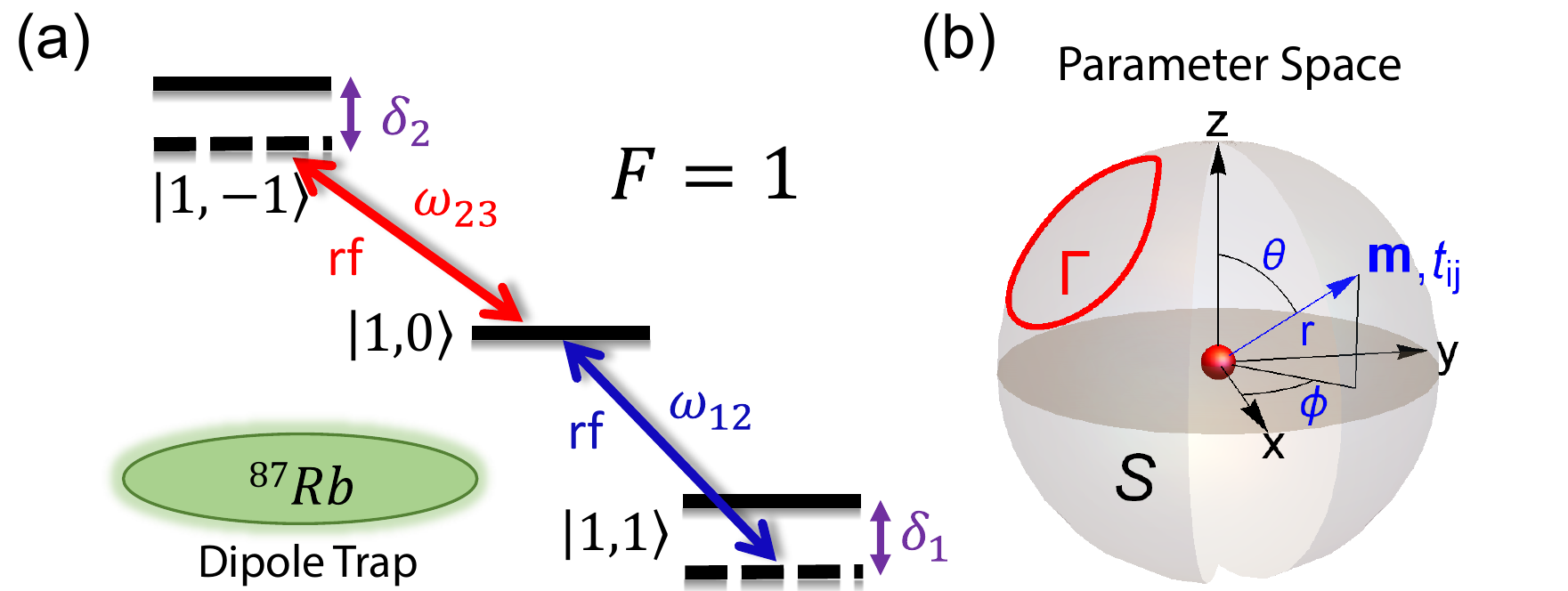}
\caption{(a) Hyperfine structure of $^{87}$Rb. The three hyperfine ground states of $|F=1\rangle $ are coupled by two independent rf fields. $\protect\omega _{12}$ and $\protect\omega _{34}$ denote coupling frequencies of the driving field. $\protect\delta _{1}$ and $\protect\delta _{2}$ are the detunings. (b) Schematics of the parameter space. The integral surface $\bm{\mathcal{S}}$ enclosing the monopoles are parameterized by spherical coordinates $(r,\protect\theta ,\protect\phi )$, with each point labeled by vector coupling $\bm{m}$ and tensor couplings $t_{ij}$. The red curve $\Gamma$ represents an arbitrary closed evolution path in the parameter space, along which a non-trivial Berry phase accumulates.}
\label{fig1}
\end{figure}

\emph{Spin-1 topological monopole}\textit{.---}We consider a $^{87}$Rb Bose-Einstein condensate (BEC) confined in an optical dipole trap. As illustrated in Fig. \ref{fig1}(a), the three hyperfine ground levels $|F=1,m_{F}\rangle $ ($m_{F}=\pm 1,0$) are coupled using two rf fields, $\Omega _{12}\cos (\omega _{12}t+\phi _{12})$ and $\Omega _{23}\cos(\omega _{23}t+\phi _{23})$, where $\Omega _{ij}$, $\omega _{ij}$, and $\phi_{ij}$ are the amplitude, frequency and phase of the driving field that couples $i$-th and $j$-th states ($|1\rangle =|1,1\rangle $, $|2\rangle=|1,0\rangle $, $|3\rangle =|1,-1\rangle $). The driving frequency are chosen as $\hbar \omega _{12}=E_{2}-E_{1}+\delta _{1}$, $\hbar \omega_{23}=E_{3}-E_{2}-\delta _{2}$, with $\delta _{1}$ and $\delta _{2}$ the detunings. The effective spin Hamiltonian in the rotating frame is
\begin{equation}
H=\left(
\begin{array}{ccc}
\delta _{1} & \Omega _{12}e^{i\phi _{12}} & 0 \\
\Omega _{12}e^{-i\phi _{12}} & 0 & \Omega _{23}e^{i\phi _{23}} \\
0 & \Omega _{23}e^{-i\phi _{23}} & \delta _{2}
\end{array}
\right) ,
\end{equation}
which can be further represented as
\begin{equation}
H=\bm{m}\cdot \bm{F}+t_{zz}N_{zz}+t_{xz}N_{xz}+t_{yz}N_{yz},  \label{ham}
\end{equation}
in terms of the spin-1 operators. Here $\bm{F}=(F_{x},F_{y},F_{z})$ is the spin-vector of $F=1$, and $N_{ij}=(F_{i}F_{j}+F_{j}F_{i})/2-\delta _{ij}\bm{F}^{2}/3$ ($i,j=x,y,z$) denote the rank-2 spin quadrupole tensors. The six coupling parameters $m_{x}=\left( \Omega _{12}\cos \phi _{12}+\Omega_{23}\cos \phi _{23}\right) /\sqrt{2}$, $m_{y}=\left( -\Omega _{12}\sin \phi_{12}-\Omega _{23}\sin \phi _{23}\right) /\sqrt{2}$, $m_{z}=\left( \delta_{1}-\delta _{2}\right) /2$, $t_{zz}=\left( \delta _{1}+\delta _{2}\right)/2 $, $t_{xz}=\sqrt{2}(\Omega _{12}\cos \phi _{12}-\Omega _{23}\cos \phi_{23})$, and $t_{yz}=\sqrt{2}(\Omega _{12}\sin \phi _{12}-\Omega _{23}\sin\phi _{23})$ can be tuned independently by varying six parameters $\Omega_{12}$, $\Omega _{23}$, $\phi _{12}$, $\phi _{23}$, $\delta _{1}$, and $\delta _{2}$ for the two rf fields in experiments. The effective fields $m_{i}$ and $t_{ij}$ couple with both spin-vectors and spin-tensors.

These six parameters form a six-dimensional parameter space, which can host exotic topological phases that may be challenging to realize in solid-state materials. By choosing suitable parameters, we can restrict the parameter space to lower dimensions for studying various topological states in 2D or 3D. For instance, by choosing $\Omega _{12}=\Omega _{23}=\Omega _{0}$, $\phi_{12}=\phi _{23}=\varphi $, we obtain $m_{x}=\Omega _{0}\cos \varphi $, $m_{y}=-\Omega _{0}\sin \varphi $, yielding a Hamiltonian similar to 2D Rashba spin-orbit coupling $k_{x}F_{x}-k_{y}F_{y}$ for spin-1 systems.

In this paper, we focus on spin-1 topological monopoles in a 3D parameter space, with six parameters $m_{i}$, $t_{ij}$ parameterized by the 3D spherical coordinates $\left(r,\theta ,\phi \right) $ (with $0\leq r$, $0\leq\theta\leq\pi$, $0\leq\phi<2\pi$). All $m_{i}$ and $t_{ij}$ vanish at the original point $r=0$, where the energy level becomes triply-degenerate. The topological monopoles manifest themselves by the Berry curvature in this parameter space, with its charge determined by the integral of Berry curvature over a closed surface $\bm{\mathcal{S}}$, as
shown in Fig. \ref{fig1}(b). For convenience, we choose the integral surface as a sphere. By tuning the coupling coefficients $\bm{m}$ and $t_{ij}$, the Berry curvature over the sphere changes and three types of topological monopoles with $\mathcal{C}=\pm 2,\pm 1,0$ emerge for spin-1 systems, which are different from the one-type Dirac monopole of spin-1/2 systems with $\mathcal{C}=\pm 1$.

\textit{I}) $\mathcal{C}=2$ monopole from a Hamiltonian $H=\mathbf{m}\cdot\mathbf{F}$ with $\mathbf{m}=r(\sin \theta \cos \phi ,\sin \theta \sin \phi,\cos \theta )$, which describes a spin-1 atom in an effective magnetic field $\mathbf{m}$ emanating from the monopole. There is no coupling with any spin tensors. The Hamiltonian can be realized using $\Omega _{12}=\Omega_{23}=r\sin \theta /\sqrt{2}$, $\delta _{1}=-\delta _{2}=r\cos \theta $, and $\phi _{12}=\phi _{23}=-\phi $.

\textit{II}) $\mathcal{C}=1$ monopole induced by an additional spin-tensor $N_{zz}$, i.e., $H=\mathbf{m}\cdot \mathbf{F}+\alpha m_{z}F_{z}^{2}$. The Hamiltonian can be realized using the same $\Omega _{ij}$, $\phi _{ij}$ as that in \textit{I}), but with different $\delta _{1}=r(\alpha +1)\cos \theta$, $\delta _{2}=r(\alpha -1)\cos \theta $. When $|\alpha |<1$, the system is adiabatically connected to the monopole with $\mathcal{C}=2$ in \textit{I}). While when $\alpha >1$, $\mathcal{C}=1$. $\alpha =1$ is a topological phase transition point with level crossing along the north pole of $\bm{\mathcal{S}}$, i.e., $\theta =0$.

\textit{III}) $\mathcal{C}=0$ monopole induced by spin-tensor $N_{xz}$ or $N_{yz}$ (which is similar), i.e., $H=\mathbf{m}\cdot \mathbf{F+}\beta m_{x}N_{xz}$. The Hamiltonian can be realized by taking $\delta _{1}=-\delta_{2}=r\cos \theta $, $\phi _{12}=-\arctan \frac{\tan \phi }{1+\beta /2}$, $\phi _{23}=-\arctan \frac{\tan \phi }{1-\beta /2}$, $\Omega _{12}=r\sin\theta \sqrt{\lbrack (1+\beta /2)^{2}\cos ^{2}\phi +\sin ^{2}\varphi ]/2}$, and $\Omega _{23}=r\sin \theta \sqrt{\lbrack (1-\beta /2)^{2}\cos ^{2}\phi+\sin ^{2}\phi ]/2}$. For $|\beta |<2$, $\mathcal{C}=2$, while for $|\beta|>2$, $\mathcal{C}=0$. $\beta =2$ is the transition point, with level crossings at $\theta=\pi/4$, $\phi=0$ or $\pi$.

\begin{figure}[t]
\centering
\includegraphics[width=3.3in]{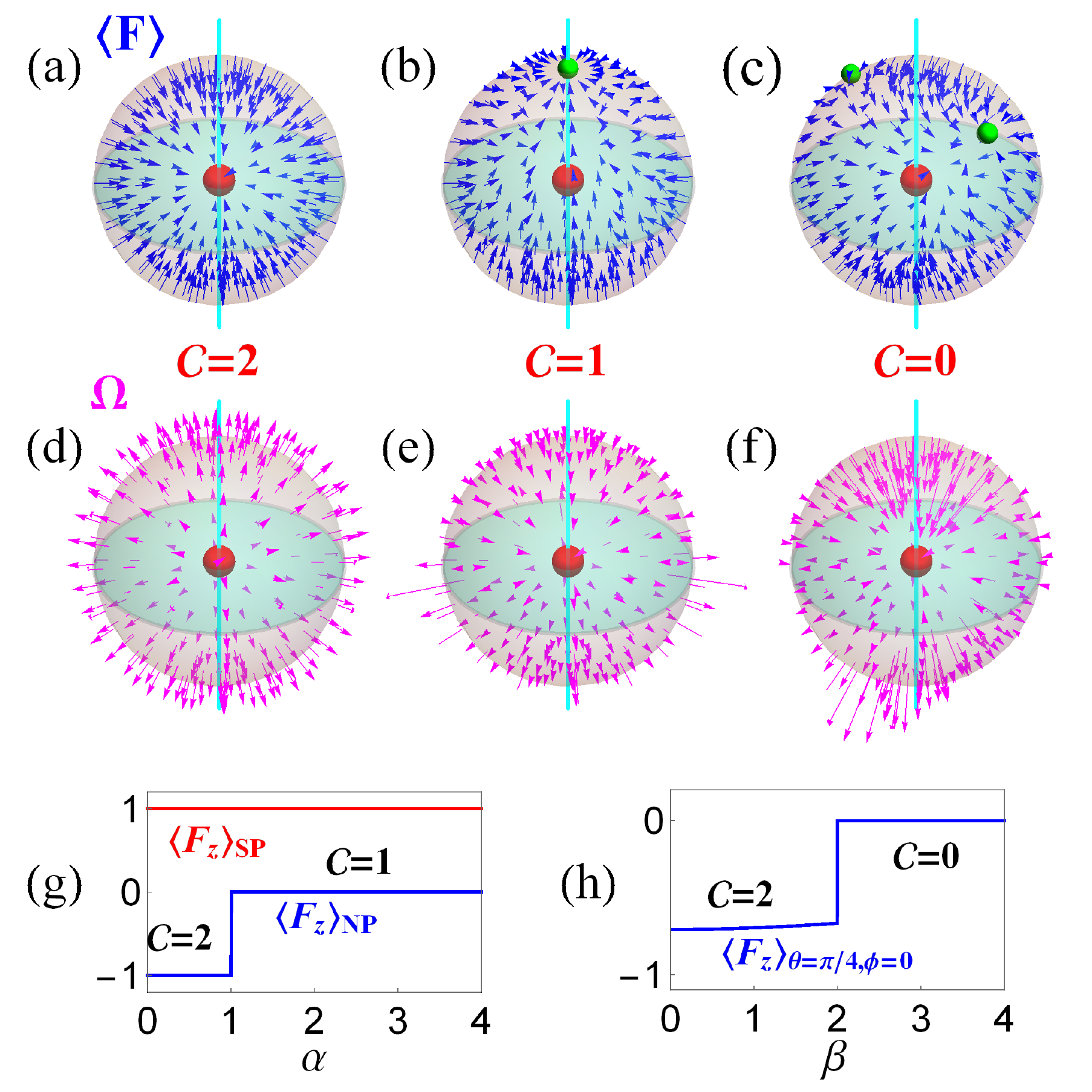}
\caption{Sketch of different types of spin-1 monopoles. The monopoles are at the origin, with the spin polarization $\langle \bm{F}\rangle $ ((a) to (c)) and Berry curvature $\bm{\Omega}$ ((d) to (f)) distributed on the surface $\bm{\mathcal{S}}$. (a)(d) are for $\mathcal{C}=2$ monopole; (b)(d) are for $\mathcal{C}=1$ monopole; and (c)(f) are for $\mathcal{C}=0$ monopole. The green dots on the sphere represent the emergent spin vortices. (g) shows the z-polarization $\langle F_{z}\rangle $ at south pole (SP) and north pole (NP) with respect to $\protect\alpha$; (h) shows the z-polarization at $\protect\theta =\protect\pi /4$, $\protect\phi =0$.}
\label{fig2}
\end{figure}

\emph{Spin texture and spin vortex}\textit{.---}The three types of spin-1 monopoles with different Chern numbers possess distinct configurations of spin textures. In Figs. \ref{fig2}(a) to (c), we illustrate the spin polarization $\langle\bm{F}\rangle\equiv\langle \psi |\bm{F}|\psi \rangle $ on the surface $\bm{\mathcal{S}}$. Similarly, the induced effective magnetic field (i.e., Berry curvature $\bm{\Omega}$) from the magnetic monopole exhibit different structures as shown in Figs. \ref{fig2}(d) to (f). For type-I monopoles with $H=\mathbf{m}\cdot \mathbf{F}$ and $\mathcal{C}=2$ [Fig. \ref{fig2} (a)(d)],
$\bm{\Omega}=-\langle \bm{F}\rangle /r^{3}=\bm{m}/r^{3}$. Because only spin-vectors appear in the Hamiltonian, $|\langle \bm{F}\rangle |=1$. The Berry curvature field is antiparallel to the spin polarization and distributes uniformly on the sphere $\bm{\mathcal{S}}$, emanating from the monopole charge located at the center.

The inclusion of spin-tensors leads to non-uniform distributions for both $\langle \bm{F}\rangle $ and $\bm{\Omega}$. Note that unlike spin-1/2 case, $|\langle \bm{F}\rangle|$ is not quantized (to $1$) for general spin-1 systems. The topological transitions between different types of monopoles occur due to level crossings, which are accompanied by the creation or annihilation of spin \textquotedblleft vortex\textquotedblright\ structure at the level-crossing points. At the core of the vortex, $\langle \bm{F}\rangle =0$, and along a small encircling loop, the direction of spin polarization winds up $2\pi$ angle. For $\mathcal{C}=1$ monopole, the created vortex resides at the north pole $\theta =0$, as shown in Fig. \ref{fig2}(b). From a perturbation analysis (up to linear term of $\delta \theta $), the wave function near the north pole is given by $|\psi \rangle =(0,\frac{\delta \theta e^{-i\phi }}{\sqrt{2}(\alpha-1)},1)^{T}$ for $\alpha <1$ and $|\psi \rangle =(-\frac{\delta \theta e^{-i\phi }}{\sqrt{2}(1+\alpha)},1,\frac{\delta \theta e^{i\phi }}{\sqrt{2}(1-\alpha)})^{T}$ for $\alpha >1$, yielding the spin polarization $\langle \bm{F}\rangle =(\frac{\delta \theta\cos \phi }{\alpha-1},\frac{\delta \theta \sin \phi }{\alpha -1},-1)\approx(0,0,-1)$ for $\alpha <1$ and $\langle \bm{F}\rangle =\frac{2\alpha\delta \theta }{1-\alpha^2}(\cos \phi ,\sin \phi ,0)$ for $\alpha >1$. It is clear for $\alpha >1$, a spin vortex is created at the north pole.

For $\mathcal{C}=0$ monopole, two spin vortices are created which are located at $\theta =\pi /4$, $\phi =0$ or $\pi $ respectively, as shown in Fig. \ref{fig2}(c). We choose the vortex at $\theta =\pi /4$, $\phi =0$ as an example. Using perturbation theory, the spin polarization near the vortex core for $\beta >2$ is $\langle \bm{F}\rangle =\frac{4\beta}{4-\beta^{2}}(\delta \theta,\delta \phi/2,-\delta \theta )$. As $\langle \bm{F}\rangle \cdot \bm{m}_{\theta =\pi /4,\phi =0}=0$, the spin polarization lies on the sphere $\bm{\mathcal{S}}$ and winds $2\pi$ along a closed path around the vortex core.

These severe changes of spin textures $\langle \bm{F}\rangle $, in particular the emergence of spin vortices, can be directly measured in experiments and thus be used to determine different types of monopoles and their phase transitions. Physically, we can interpret topological phase transitions as the transfer of singularity from the monopole charges to the emergent spin vortices. Experimentally, the transition between $\mathcal{C}=2$ and $\mathcal{C}=1$ monopoles can be determined by directly measuring the spin polarization along the $z$-direction:8 $\langle F_{z}\rangle =\frac{N_{1}-N_{-1}}{N_{1}+N_{0}+N_{-1}}$ on the two poles of $\bm{\mathcal{S}}$, as shown in Fig. \ref{fig2}(g). Here $N_{1}$, $N_{0}$, and $N_{-1}$ denote the populations of corresponding hyperfine states. For a $\mathcal{C}=2$ monopole, $\langle F_{z}\rangle _{SP}=-\langle F_{z}\rangle _{NP}=1$; While for $\mathcal{C}=1$ case, $\langle F_{z}\rangle _{SP}=1$, $\langle F_{z}\rangle _{NP}=0$. For a $\mathcal{C}=0$ monopole as shown in Fig. \ref{fig2}(c) and (f), the spin polarization and Berry curvature field are mainly distributed near the two poles, with opposite directions for the Berry curvature field dictated by its zero monopole charge. Similarly, the transition between $\mathcal{C}=2$ and $\mathcal{C}=0$ monopoles can be experimentally determined by measuring the spin polarization at $\theta =\pi /4$ and $\phi =0$ as illustrated in Fig. \ref{fig2} (h). For $\mathcal{C}=2$ case, $\langle F_{x}\rangle =\langle F_{z}\rangle =-4/\sqrt{32+\beta ^{2}}$, $\langle F_{y}\rangle =0$, while for $\mathcal{C}=0$, $\langle \bm{F}\rangle=\bm{0}$.

\emph{MSR for topological monopoles}\textit{.---}The emergence of different types of spin-1 monopoles and their topological phase transitions can also be intuitively understood and visualized by utilizing a geometric method---MSR \cite{majorana1}, which projects states in high-dimensional Hilbert space to a few points, named Majorana stars, on the Bloch sphere. Each Majorana star corresponds to an individual spin-1/2 state. The physical quantities are then reflected on the geometrical properties of these Majorana stars.

In a spin-1/2 system, any state can be written as the superposition $|\psi\rangle =\cos \frac{\xi }{2}|\uparrow \rangle +e^{i\eta }\sin \frac{\xi }{2}|\downarrow \rangle $, with $0\leq \xi \leq \pi $ and $0\leq \eta < 2\pi $. The state $|\psi \rangle $ is in exact correspondence with a point $\bm{u}=(\sin \xi \cos \eta ,\sin \xi \sin \eta ,\cos \xi )$ on the Bloch sphere, i.e., each spin-1/2 state is represented by one Majorana star. Here $\xi$ and $\eta $ denote the colatitude and longitude in the spherical coordinate.
For an arbitrary three component state $|\psi \rangle =f_{-1}|1,-1\rangle+f_{0}|1,0\rangle +f_{1}|1,1\rangle $, we can use the Schwinger boson theory \cite{schwinger} to rewrite the spin-1 basis by the creation and annihilation operators of two-mode bosons $a^{\dag }$, $a$, and $b^{\dag }$, $b$: $|1,m\rangle =\frac{(a^{\dag })^{1+m}(b^{\dag })^{1-m}}{(1+m)!(1-m)!}|\emptyset \rangle $. The state $|\psi \rangle $ is then factorized as
\begin{equation}
|\psi \rangle =\frac{1}{\mathcal{N}}\prod_{j=1}^{2}(\cos \frac{\xi _{j}}{2}a^{\dag}+\sin \frac{\xi _{j}}{2}e^{i\eta _{j}}b^{\dag })|\emptyset \rangle ,\label{ms}
\end{equation}
where $\mathcal{N}$ is the normalization coefficient. Denote $y_{j}=\tan \frac{\xi_{j}}{2}e^{i\eta _{j}}$ and $a^{\dag }|\emptyset \rangle =|\uparrow \rangle $, $b^{\dag }|\emptyset \rangle =|\downarrow \rangle $, then $y_{j}$ satisfies $\sum_{j=0}^{2}\frac{(-1)^{j}f_{1-j}}{\sqrt{(2-j)!j!}}y^{2-j}=0$. From Eq. (\ref{ms}), it is obvious that any spin-1 state can be characterized by two individual Majorana stars $\bm{u}_{j}=(\sin \xi_{j}\cos \eta _{j},\sin \xi _{j}\sin \eta _{j},\cos \xi _{j})$ ($j=1,2$) on the Bloch sphere.

\begin{figure}[t]
\centering
\includegraphics[width=3.3in]{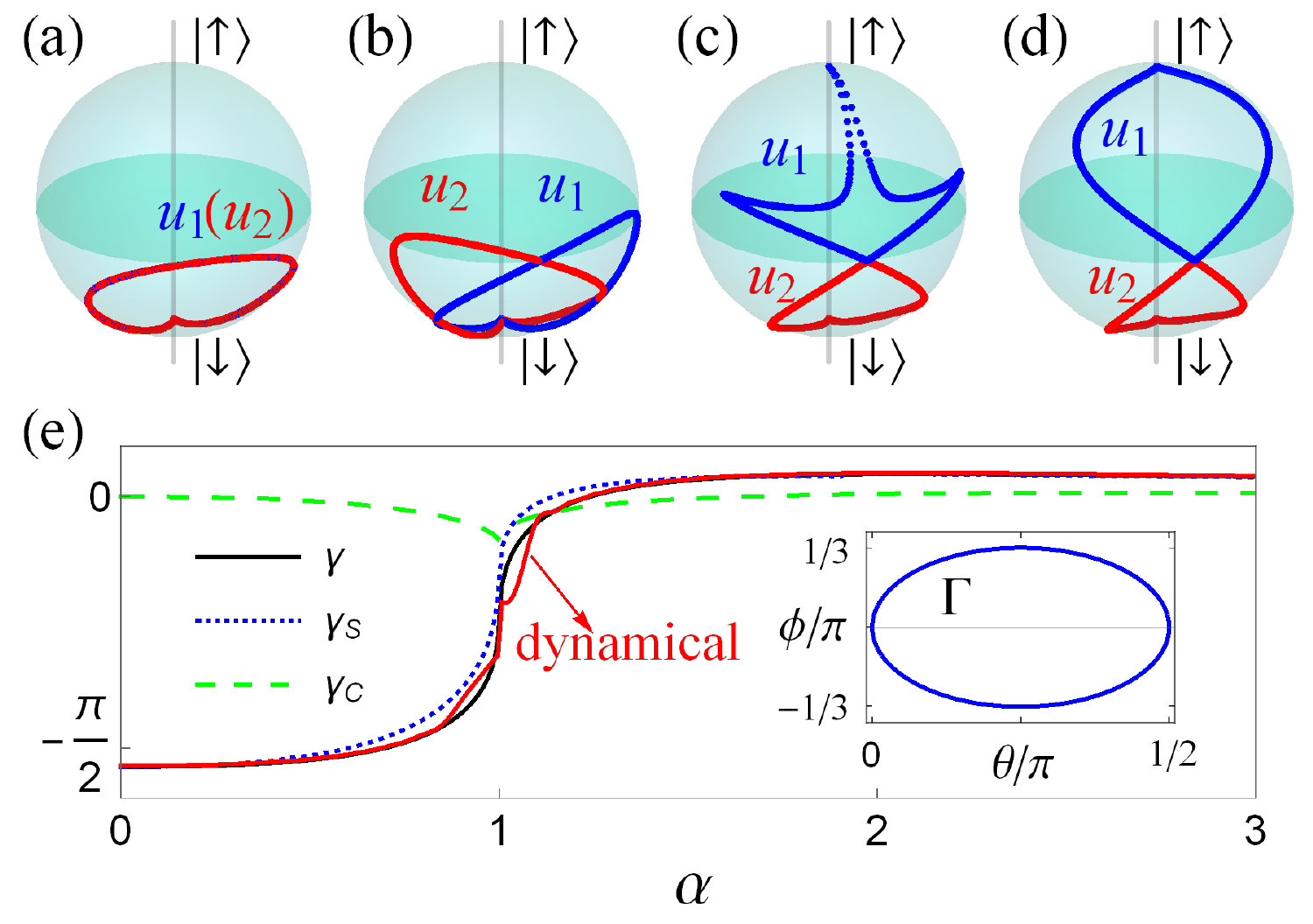}
\caption{Trajectories of two Majorana stars $\bm{u}_{1}$ (blue) and $\bm{u}_{2}$ (red) on the Bloch sphere for different types of spin-1 monopoles with respect to evolution path $\Gamma (t)$. From (a) to (d), $\protect\alpha =0$, $0.5$, $1.001$, $2$. (e) Berry phase of the ground state associated with path $\Gamma (t)$. Black, blue dotted, and green dashed lines show the total Berry phase $\protect\gamma $, the solid angle part $\protect\gamma _{s}$, and the correlation part $\protect\gamma _{c}$. The red line is extracted from the dynamical protocol. The inset illustrates the evolution path $\Gamma (t)$ in the parameter space.}
\label{fig3}
\end{figure}

The trajectories of two Majorana stars on the Bloch sphere can be used to visualize different types of topological monopoles and their topological phase transitions. Here we take type-II monopoles as an example and consider a closed evolution path
\begin{equation}
\Gamma (t):~~~\theta (t)=\frac{\pi }{4}\cos \frac{2\pi t}{T}+\frac{\pi }{4},~~~\phi (t)=\frac{\pi }{3}\sin \frac{2\pi t}{T}.
\end{equation}
in the parameter space with $\Gamma (t)=\Gamma (t+T)$. In Figs. \ref{fig3}(a-d), the trajectories of two Majorana stars for the ground states at four typical $\alpha $'s are drawn. At $\alpha =0$, two Majorana stars $\bm{u}_{1}$ and $\bm{u}_{2}$ coincide with each other, starting and ending at the south pole of the Bloch sphere [Fig. \ref{fig3}(a)]. This corresponds to a spin-1 in a magnetic field $\mathbf{m}$ of parameter space with eigenstate $e^{iF_{y}\theta }e^{iF_{z}\phi }|1,-1\rangle $. By increasing $\alpha $, two Majorana stars start to separate, as shown in Fig. \ref{fig3}(b), while still share the same starting and ending points at the south pole. Further increasing $\alpha $ till $\alpha =1$, the trajectories \textquotedblleft explode\textquotedblright\ on the Bloch sphere, accompanied by a sudden change of their topology at $\alpha =1$ [Fig. \ref{fig3}(c)]. After the transition point, while one of Majorana stars is still bonded to the south pole, the trajectory of the other one now starts and ends at the north pole as shown in Fig. \ref{fig3}(d). Two Majorana stars only share one touching point on the equator.

Similar analysis can be performed for the type-III monopoles and their phase transitions \cite{supp}. We note that while the configurations of two Majorana stars rely on the selection of evolution path, the change of the topology of the trajectories is always accompanied by the phase transition between different types of monopoles, revealing their distinct geometric and topological properties.

\emph{Berry phase and Berry curvature}\textit{.---}In an adiabatic evolution over a course of cycle, Berry phase, which results from the geometric properties of the underlying Hamiltonian, can be represented as the integral of gauge potential: $\gamma =\oint_{\Gamma }\bm{A}\cdot d\bm{R}\equiv i\oint_{\Gamma}\langle \psi |\nabla _{\bm{R}}|\psi \rangle \cdot d\bm{R}$. For a spin-1/2 system, $\gamma $ is simply the solid angle subtended by the trajectories of the Majorana star. While for a spin-1 system, any quantum state is represented by two Majorana stars. Fortunately, the Berry phase accumulated along a closed path can be elegantly formulated as \cite{majorana2,majorana3,majorana4,majorana5}
\begin{eqnarray}
\gamma =\gamma _{S}+\gamma _{C}= &-&\sum_{j=1}^{2}\frac{1}{2}\oint (1-\cos\xi _{j})d\eta _{j}  \notag \\
&-&\frac{1}{2}\oint \frac{(d\bm{u}_{1}-d\bm{u}_{2})\cdot (\bm{u}_{1}\times\bm{u}_{2})}{3+\bm{u}_{1}\cdot \bm{u}_{2}},  \label{bphase}
\end{eqnarray}
where the first term $\gamma _{S}$ denotes the solid angles traced out by two Majorana stars and the second term $\gamma _{C}$ describes their correlations due to the relative motion.

In Fig. \ref{fig3}(e), we show the Berry phase along path $\Gamma (t)$ with respect to $\alpha $. At $\alpha =0$, two Majorana stars coincide, hence the correlation part vanishes ($\gamma _{C}=0$) and $\gamma$ is twice the solid angle subtended by each Majorana star. By increasing $\alpha $, the trajectories of two Majorana stars separate and their correlation $\gamma_{C}$ becomes nonzero. At $\alpha =1$, the three phases exhibit discontinuities due to the change of topology of two Majorana stars. After the
transition point, $\gamma $ tends to zero, in consistent with that two Majorana stars are bonded to different poles on the Bloch sphere [Fig. \ref{fig3}(d)].

The Berry phase $\gamma $ can also be obtained from the Berry curvature $\bm{\Omega}$ by $\gamma =\iint_{S_{\Gamma }}\bm{\Omega}\cdot d\bm{S}$, with the boundary $\partial S_{\Gamma }=\Gamma $. If the integral is performed on a closed 2D manifold, it gives the monopole charge. Using MSR, the Berry curvature takes the following form \cite{MSRcurv}:
\begin{eqnarray}
\bm{\Omega}_{\alpha \beta } &=&-\frac{2}{(3+\bm{u}_{1}\cdot \bm{u}_{2})^{2}}{\large [2}\sum\nolimits_{i=1}^{2}{\large \bm{u}_{i}\cdot (\partial _{\alpha}\bm{u}_{i}\times \partial _{\beta }\bm{u}_{i})}  \notag \\
&&{\large +(\bm{u}_{1}+\bm{u}_{2})\cdot (\partial _{\alpha }\bm{u}_{1}\times\partial _{\beta }\bm{u}_{2}+\bm{u}_1\leftrightarrow\bm{u}_2)]}.  \label{BC}
\end{eqnarray}
For type-I monopoles without spin-tensors, $\bm{u}_{1}=\bm{u}_{2}\equiv\bm{u}$, and Eq. (\ref{BC}) reduces to $\bm{\Omega}_{\alpha \beta }=-\bm{u}\cdot (\partial _{\alpha }\bm{u}\times \partial _{\beta }\bm{u})$, indicating that the monopole charge is the winding number of two Majorana stars on the Bloch sphere. The spin-tensors deform the configurations of Majorana stars, hence change the topological charge of the monopole.

For the integral sphere $\bm{\mathcal{S}}$ in the parameter space in Fig. \ref{fig1}(b), the topological charge can be written as
\begin{equation}
\mathcal{C}=\frac{1}{2\pi }\int_{0}^{\pi }d\theta \int_{0}^{2\pi }d\phi~\Omega _{\theta \phi },  \label{topoinvariant}
\end{equation}
where $\Omega _{\theta \phi }=\text{Im}{\large [\langle \frac{\partial \psi}{\partial \theta }|\frac{\partial \psi }{\partial \phi }\rangle -\langle\frac{\partial \psi }{\partial \phi }|\frac{\partial \psi }{\partial \theta }\rangle ]}$ is the Berry curvature in spherical coordinates.

\textit{Experimental detection of monopole charge}\textit{.---}The spin textures and the emergence of spin vortices in the previous discussions provide a simple experimental tool for distinguishing different types of monopoles and their phase transitions. To measure the Berry phase and determine the monopole charge, we need to measure the Berry curvature on each point in the parameter space, which can be done using non-adiabatic effect \cite{pnas} during the ramping of certain related parameter $\lambda $. The
non-adiabaticity leads to the deflections of quantum trajectories that are proportional to the Berry curvature in parameter space, analogous to a charged particle moving in a magnetic field deflected by Lorentz force. Formally, the deflection is described by a generalized force $\bm{M}_{\mu }=-\langle \partial _{\mu}H\rangle $, and related to Berry curvature through
\begin{equation}
\mathbf{M}_{\mu }=-\langle \psi _{0}(\bm{R})|\partial _{\mu }H|\psi _{0}(\bm{R})\rangle +\mathbf{v}_{\lambda }\times \mathbf{\Omega }_{\lambda \mu}+O(\bm{v}_{\lambda }^{2}).  \label{GF}
\end{equation}
using linear response theory \cite{pnas}. Here $\psi _{0}(\bm{R})$ is the spontaneous eigenstate at $\bm{R}$. The last term denotes higher-order corrections. $\bm{v}_{\lambda }=\frac{d\lambda }{dt}$ is the ramping velocity of $\lambda$. It is easy to verify that the contribution from the first term is zero for the integral on a closed surface. Considering that adiabaticity is usually hard to achieve in realistic laboratory condition, this relation has the advantage of needing only a moderately slow change of parameters with dominating linear terms.
\begin{figure}[t]
\centering
\includegraphics[width=3.3in]{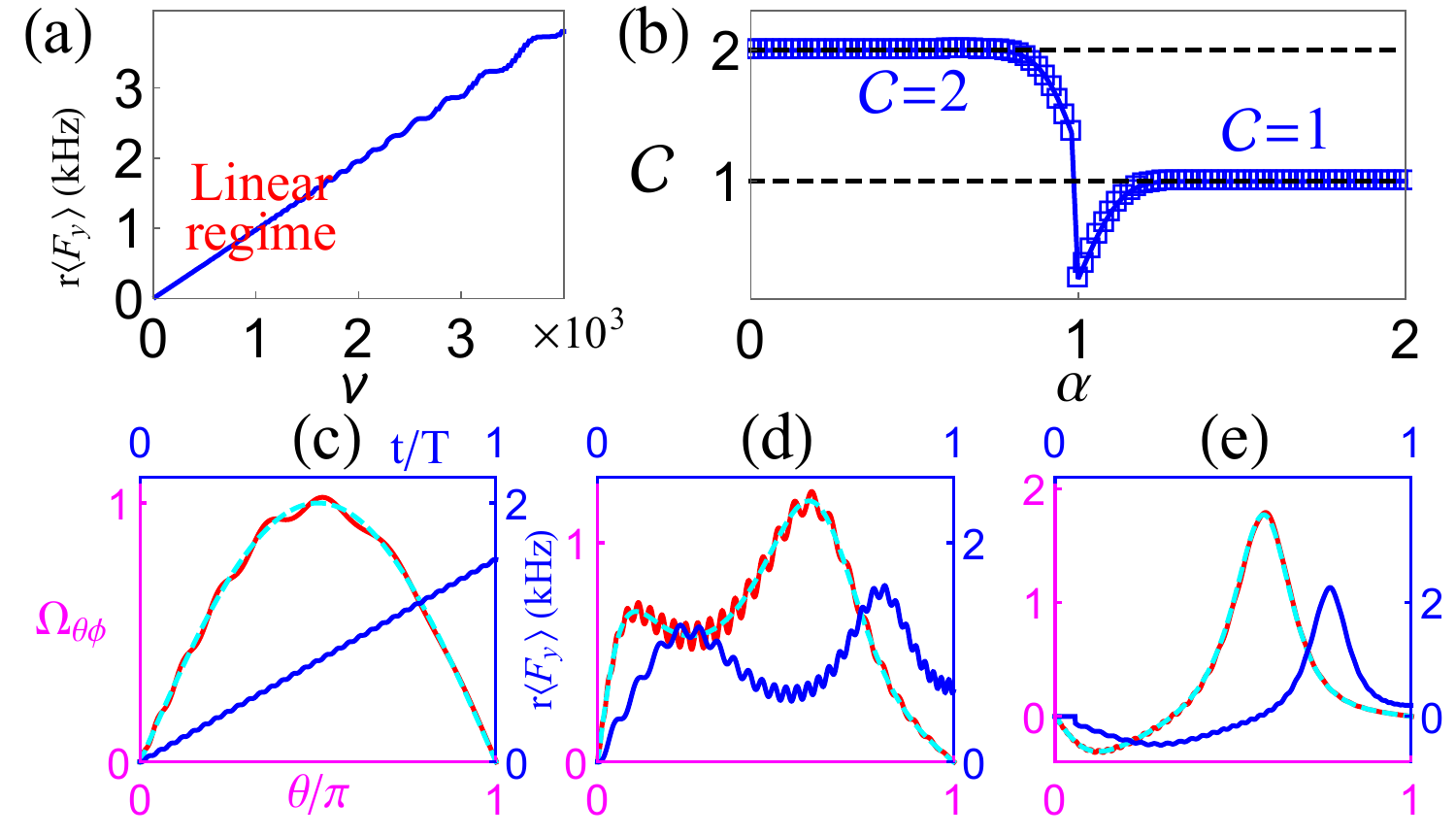}
\caption{Berry curvature and monopole charge from dynamical effects. (a) Plot of $r\langle F_{y}\rangle $ with respect to $v$. The slope in the linear regime for small $v$ gives $\Omega _{\protect\theta \protect\phi }$. (b) Monopole charge with respect to $\protect\alpha $. (c), (d), (e) show $\Omega _{\protect\theta \protect\phi }$ and $\langle F_{y}\rangle $ (blue solid line) for $\protect\alpha =0$, $0.5$, and $1.8$, respectively. The red lines are calculated from the dynamical effects, and the cyan dashed lines indicate theoretical values. $r=16\protect\pi \times $ kHz, $T=4$ ms, which are in the linear regime.}
\label{fig4}
\end{figure}

For type-I and type-II monopoles, the Hamiltonian is cylindrically invariant. Accordingly, the Berry curvature must be cylindrically symmetric $\Omega _{\theta ,\phi }=\Omega _{\theta ,\phi =0}$. The force along the longitude direction is given by $M_{\phi }=-\langle \partial _{\phi }H\rangle=r\sin \theta \sin \phi \langle F_{x}\rangle -r\sin \theta \cos \phi \langle F_{y}\rangle $. Hence $\mathcal{C}=\int_{0}^{\pi }\Omega _{\theta ,\phi=0}d\theta =\int dt\sin \theta ~r\langle F_{y}\rangle $. We choose a smooth
evolution path: $\theta =v^{2}t^{2}/2\pi $ with $v_{\theta }=v^{2}t/\pi $, which is adiabatic at $t=0$ and at $t=\pi /v$, $v_{\theta }=v$.

In Fig. \ref{fig4}(a) we show the dependence of $\Omega _{\theta \phi }$ at $t=\pi /v$ on evolution speed $v$. It is clear for small $v$, the dynamical evolution lies in a linear regime, where the higher-order corrections are negligible. The Berry curvature $\Omega _{\theta \phi }$ can then be extracted from the slope of the curve, in consistent with theoretical value $\Omega_{\theta \phi }=\sin \theta|_{t=\pi/v} =1$ at $\alpha =0$. Now we constrain the discussions in this linear regime. In Fig. \ref{fig2}(e), we plot the Berry phase calculated from the dynamical effect, which agrees quite well with the theoretical values obtained from Eq. (\ref{bphase}). The integrated monopole charge $\mathcal{C}$ is shown in Fig. \ref{fig4}(b). For $\alpha <1$, $\mathcal{C}$ is quantized to $2$ while for $\alpha >1$, $\mathcal{C}$ is quantized to $1$. The system undergoes a topological phase transition at $\alpha =1$, characterized by the change of monopole charge. Note that near the phase transition point, $\mathcal{C}$ is not precisely quantized due to the
small energy gap in the evolution process.

The time-dependent magnetization $\langle F_{y}\rangle $ and the extracted Berry curvature for different $\alpha $ are shown in Figs. \ref{fig4}(c)-(e). At $\alpha =0$, $\langle F_{y}\rangle $ is linearly dependent on $t$ with small oscillations from dynamical effects [Fig. \ref{fig4}(c)]. The extracted Berry curvature is in consistent with theoretical value $\Omega_{\theta \phi }=\sin \theta $. With increasing $\alpha $, both $\Omega_{\theta \phi }$ and $\langle F_{y}\rangle $ exhibit two peaks, accompanied by larger oscillations (energy gap decreases by increasing $\alpha $) as shown in Fig. \ref{fig4}(d). $\mathcal{C}$ is still quantized to $2$. Further increasing $\alpha $ to the transition point, the left peak of $\Omega _{\theta \phi }$ moves towards the boundary $\theta =0$. After that,
a negative peak emerges near the same boundary as shown in Fig. \ref{fig4}(e), accompanied by a sudden change of $\mathcal{C}$.
\begin{figure}[t]
\centering
\includegraphics[width=3.3in]{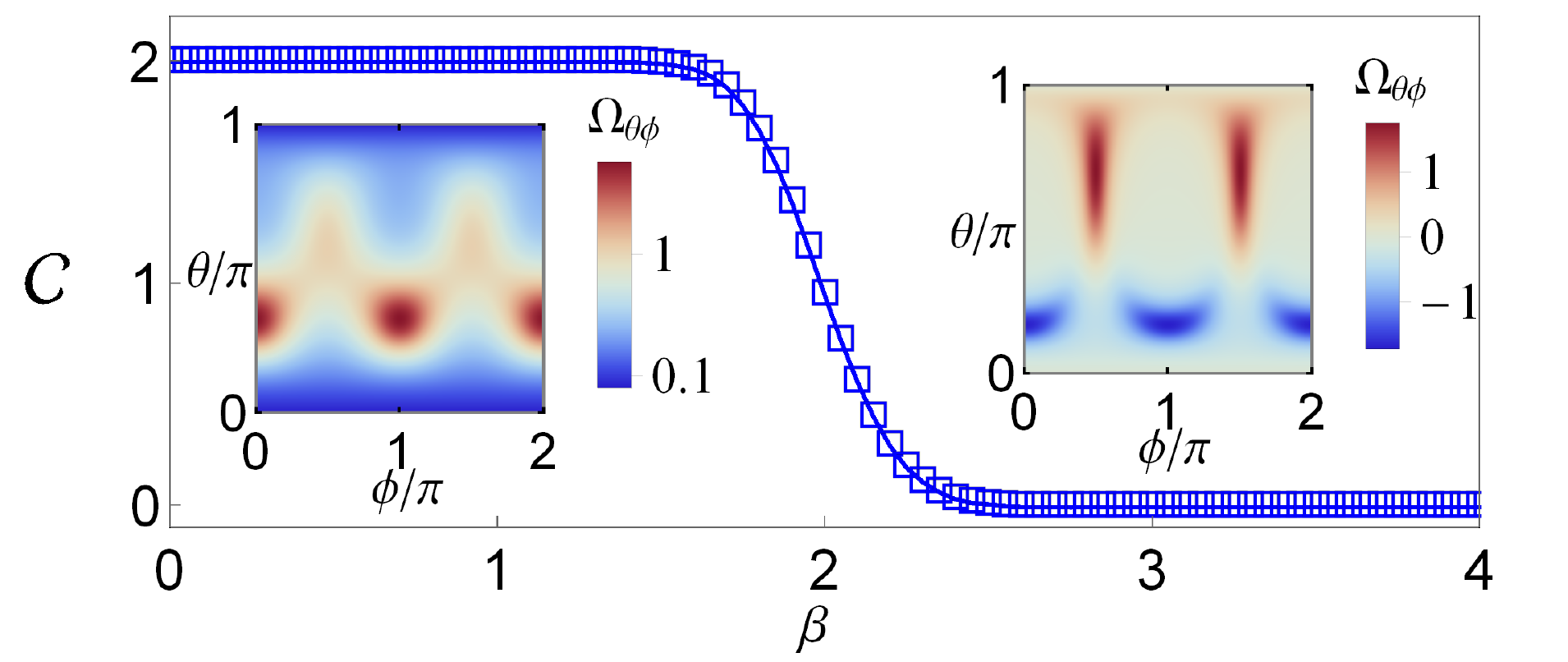}
\caption{Monopole charge extracted from dynamical effects for type-III. The main figure shows $\mathcal{C}$ with respect to $\protect\beta $. The insets show the Berry curvatures for $\protect\beta =0.5$ and $\protect\beta =4$. $r=16\protect\pi \times $ kHz, $T=4$ ms.}\label{fig5}
\end{figure}

For type-III monopoles, we have $M_{\phi }=-\langle \partial _{\phi}H\rangle =r\sin \theta \sin \phi \big[\langle F_{x}\rangle +\beta \langle N_{xz}\rangle \big]-r\sin \theta \cos \phi \langle F_{y}\rangle $. The monopole charge can then be extracted by measuring spin vectors $\langle F_{x}\rangle $, $\langle F_{y}\rangle $ and spin tensor $\langle N_{xz}\rangle $ for different ($\theta $,$\phi $) in the parameter space. The main results are summarized in Fig. \ref{fig5}. We can clearly see $\beta =2$ is a phase transition point, with $\mathcal{C}$ changing from $2$ to $0$. The Berry curvature shows different behaviors for two phases. At $\beta <2$, four positive peaks appear at $\phi =n\pi /2$. While across the transition point, the peaks at $\phi =0,\pi $ turn into negative peaks,
cancelling the Berry curvature field in other regions. The integrated Berry curvature then gives $\mathcal{C}=0$ for $\beta >2$.

\textit{Conclusion}\textit{.---}To summarize, we have demonstrated a versatile ultracold atomic platform for the generation, manipulation and observation of various topological quantum phases such as spin-1 topological monopoles in parameter space. Our proposed simple experimental system involves only two rf fields to couple three different hyperfine states of ultracold atoms, which define relevant parameter spaces, paving the way for exploring and engineering new exotic quantum matter.

\begin{acknowledgments}
This work is supported by NSF (PHY-1505496), ARO (W911NF-17-1-0128), AFOSR (FA9550-16-1-0387).
\end{acknowledgments}

\clearpage
\onecolumngrid
\appendix

\section{Supplementary Materials}

\subsection{MSR of type-III monopoles}

In this section, we show the trajectories of Majorana stars using MSR and visualize the topological phase transitions on the Bloch sphere for type-III monopoles in the main text. The evolution path is chosen as $\Gamma_{2}(t):\theta (t)=\frac{\pi }{8}\cos \frac{2\pi t}{T}+\frac{\pi }{4},\phi(t)=\frac{\pi }{4}\sin \frac{2\pi t}{T}+\frac{3\pi }{4}$.

\begin{figure}[h]
\centering
\includegraphics[width=5.5in]{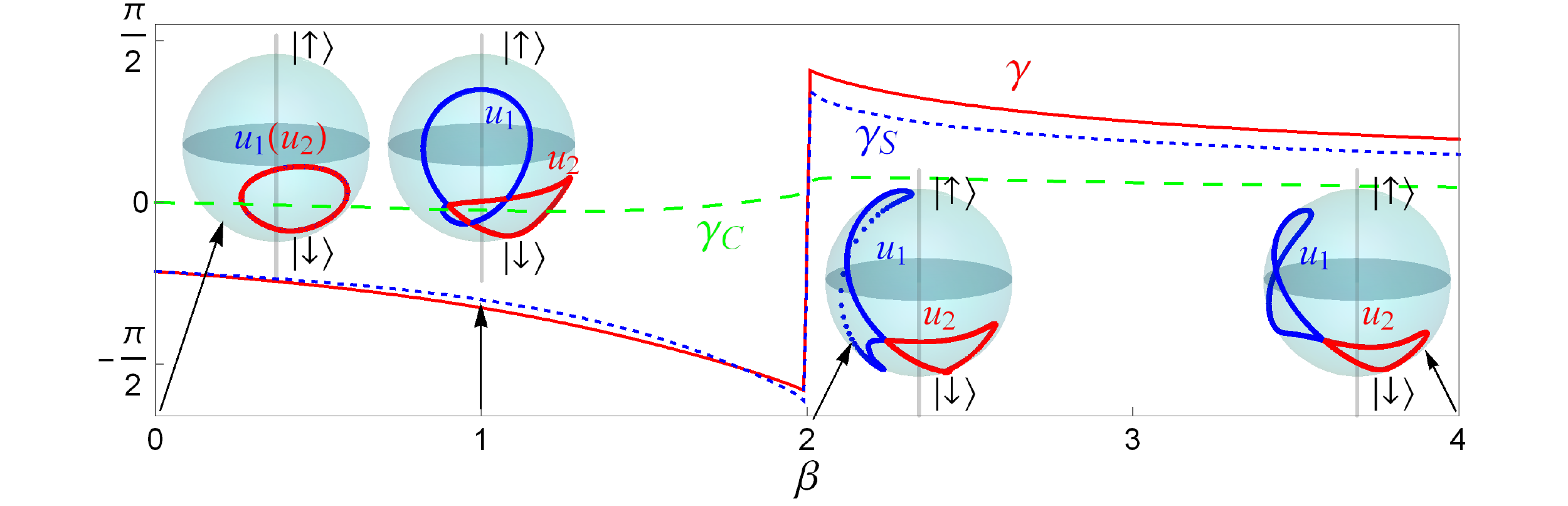}
\caption{Berry phase and MSR associated with evolution path $\Gamma _{2}(t)$ for type-III monopoles. Red solid, blue dotted and green dashed lines represent the total Berry phase $\protect\gamma $, the solid angle part $\protect\gamma _{S}$ and the correlation part $\protect\gamma _{C}$. The insets plot the trajectories of two Majorana stars $\bm{u}_{1}$ (blue) and $\bm{u}_{2}$ (red) at $\protect\beta =0$, $1$, $2.01$, $4$.}
\label{figs1}
\end{figure}

The trajectories of two Majorana stars of the ground state are shown in Fig. \ref{figs1} for four typical $\beta $. At $\beta =0$, two Majorana stars $\bm{u}_{1}$ and $\bm{u}_{2}$ coincide with each other, sharing the same curves on the Bloch sphere. Hence $\gamma _{C}=0$, and $\gamma =\gamma _{S}$. By increasing $\beta $, two Majorana stars start to separate, as shown in Fig. \ref{figs1}(b). The two trajectories share three touching points, one of which is fixed for all $\beta $. The topological phase transition occur at $\beta =2$. After the transition, the two trajectories only share one common point. Correspondingly, the Berry phases $\gamma $, $\gamma _{S}$, $\gamma _{C}$ exhibit abrupt change at the transition point.

\end{document}